\documentstyle[aps,preprint,epsfig]{revtex}
\newcommand{\bq}{\begin{equation}} \newcommand{\eq}{\end{equation}}
\newcommand{\bqn}{\begin{eqnarray}} \newcommand{\eqn}{\end{eqnarray}}
\newcommand{\nb}{\nonumber}
\newcommand{\lb}{\label}
\def\noi{\noindent}

\begin{document}

\title{Type II Fluid Solutions to Einstein's Field Equations
in N-Dimensional Spherical Spacetimes}
\author{Jaime F. Villas da Rocha }
\address{  Departamento de F\' {\i}sica Te\' orica, 
Universidade do Estado do Rio de Janeiro, 
\small\it Rua S\~ ao Francisco Xavier $524$, Maracan\~ a, $20550-013 $
Rio de Janeiro~--~RJ, Brazil}

\maketitle

\begin{abstract}

A large class of Type II fluid solutions
to Einstein field equations in N-dimensional spherical spacetimes
is found, wich includes most of the known solutions.
A family of the generalized collapsing Vaidya solutions with homothetic
self-similarity, parametrized by a constant $\lambda$, is studied,
and found that
when $\lambda$ $>$ $\lambda_c(N)$, the collapse always forms black holes,
and when $\lambda$ $<$ $\lambda_c(N)$, it always forms naked singularities,
where $\lambda_c(N)$ is function of the spacetime dimension N only.
\end{abstract}

\vspace{.4cm}

%%%%%%%%%%%%%%%%%%%%%%%%%%%%%%%%%%%%%%%%%%%%%%%%%%%%%%%%%%%%%%%%%%%%%%%

\section{Introduction}
The Vaidya solutions 
found in 1951 \cite{V1951} represent an
imploding (exploding) null dust fluid with spherical symmetry
and has been intensively studied in gravitational
collapse in General Relativity \cite{J1992}.
Papapetrou \cite{P1984} first
showed that these solutions can give rise to the formation of naked
singularities, and thus provide 
one of the earlier counterexamples 
to the cosmic censorship conjecture \cite{P1969}. Other 
generalized Vaidya solutions with 
remarkable physically properties
are, for example, the charged ones \cite{LSM1965}
and the Husain solutions
\cite{H1996} of a null fluid with a particular equation of state,
which  have been lately used
as the formation of black holes with short hair \cite{BH1997}.
Recently, Wang and Wu \cite{Wang1998} 
generalized the Vaidya solutions to a more
general case, which include most of the known solutions 
to the Einstein field equations, such as, 
the de Sitter-charged Vaidya solutions, and
the Husain solutions. More recently,  a class of the Wang-Wu solutions
was used to study the collapse of strange quark matter in a Vaidya background
\cite{HC00}.

On the outher hand, Waugh and Lake \cite{WL86}
studied the Vaidya solutions in double null coordinates
and found the explicit close form of the metric coefficients
for the linear and exponential mass functions. As a result,
the global structure of the spacetime was showed 
clearly. In particular,
they showed that for the homothetic
case, where the mass function takes the form
$m(v)$  $=$ $\lambda v$, when $\lambda > 1/16$ the collapse
forms black holes and when 
$\lambda < 1/16$ the collapse always forms naked 
singularities. 
This is quite similar to the critical phenomena in gravitational
collapse found recently by Choptuick \cite{Ch93},
but now the ``critical'' solution $\lambda = 1/16$
separates black holes from naked singularities.
To have a definite answer to this problem,
one needs to study the perturbations of the
``critical'' solutions and shows  that it indeed has only one unstable
mode. Otherwise, this ``critical'' solutions 
is not critical by definition \cite{GW86}.

In this paper, we first generalize the results obtained in
\cite{Wang1998} to N dimensional spherically symmmetric spacetimes
in Eddington-Finkelstein (radiation) coordinates,
and then study the physical properties of 
generalized Vaidya ingoing solutions with homothetic
self-similarity. In particular, we shall show that, similar the
four dimensional
case \cite{WL86}, there always exists a ``critical'' solution
$\lambda = \lambda_c(N)$, which separates the formation
of black holes from the formation
of naked singularities. It is remarkable
that  $\lambda_c(N)$ is function of the spacetime dimension N, only.

\section{ Type II fluid solutions in 
N-Dimensional Spherical Spacetimes }

Let us begin with the general spherically symmetric line element
\cite{jai99}
\bq
\lb{eq1}
ds^2= e^{\psi(v,r)}dv
\left[ f(v,r)e^{\psi(v,r)}dv +2\epsilon dr \right]
-C^2(v,r)d\Omega^{2}_{N-2},
\eq
\noindent where
$d\Omega^{2}_{N-2}$ is the line element
on the unit (N-2)-sphere, given by 
\bqn
\lb{eq1a}
d\Omega^{2}_{N-2} &=& \left(d\theta^{2}\right)^{2} +
\sin^{2}(\theta^{2})\left(d\theta^{3}\right)^{2} + 
\sin^{2}(\theta^{2})\sin^{2}(\theta^{3})\left(d\theta^{4}\right)^{2}\nb\\
& & + ... + \sin^{2}(\theta^{2})\sin^{2}(\theta^{3}) ...
\sin^{2}(\theta^{N-2})\left(d\theta^{N-1}\right)^{2}\nb\\
& = &  \sum^{N-1}_{i = 2}{\left[
\prod^{i-1}_{j =2}\sin^{2}(\theta^{j})\right]
\left(d\theta^{i}\right)^2}\eqn
and $\epsilon = \pm 1$. 
When $\epsilon = + 1$, the radial coordinate $r$
increases toward the future along a ray 
$v = Const.$, i.e., the light cone 
$v = Const.$ is expanding.  When $\epsilon = - 1$, 
the radial coordinate $r$
decreases toward the future along a ray 
$v = Const.$, and the light cone 
$v = Const.$ is contracting. 

In the
following, we shall consider the particular 
case where $\psi(v, r) = 0$
and  $\epsilon = - 1$
then the non-vanishing components of the Einstein 
tensor are given by
\bqn
\lb{eq2}
G^{0}_{0} = G^{1}_{1} & =  & 
-\frac{(N-2)}{2r^2}\left\{(N-3)[1 - f(v,r)] - rf'(v, r)\right\},  \\
G^{1}_{0} & = &- \frac{(N-2)}{2 r} \dot{f}(v, r), \\
G^{i}_{i}
&=& \frac{1}{2r^2} \left\{
r^2 f''(v, r) + (N-3)[2r f'(v,r) + \phantom{\frac{1}{r^2}}
\right.  \nb \\
& & \left. \phantom{- \frac{1}{2r^2}+} 
- (N-4)(1-f(v,r)]\right\},
\eqn
where $\{x^{\mu}\} = \{v, r, \theta^2, 
...,\theta^{N - 1} \},\; (\mu = 0, 1, 2,..N-1)$,
and
\bq
\dot{f}(v, r) \equiv \frac{\partial f(v, r)}{\partial v},
\;\;\;\;\;\;
f'(v, r) \equiv \frac{\partial f(v, r)}{\partial r}.
\eq
Then, from the Einstein field equations $G_{\mu\nu} =
\kappa T_{\mu\nu}$, we find that the corresponding EMT can be written in
the form \cite{H1996}
\bq
\lb{eq3}
T_{\mu\nu} = T^{(n)}_{\mu\nu} + T^{(m)}_{\mu\nu},
\eq
where
\bqn
\lb{eq4}
T^{(n)}_{\mu\nu} &=& \mu l_{\mu}l_{\nu},\nb\\
T^{(m)}_{\mu\nu} &=& (\rho + P) \left(l_{\mu}n_{\nu} +
l_{\nu}n_{\mu}\right) + P g_{\mu\nu},
\eqn
and
\bqn
\nb
\mu  & = & -  \frac{(N-2)}{2\kappa r} \dot{f}(v, r),  \\
\lb{eq5}
\rho & = &   \frac{(N-2)}{2 \kappa r^2}
\left\{(N-3)[1 - f(v,r)] - rf'(v, r)\right\},  
\\
\nb
P &= &
\frac{1}{2 \kappa r^2} \left\{
r^2 f''(v, r) + (N-3)[2r f'(v,r)- (N-4)(1-f(v,r))\right\},
\eqn
with $l_{\mu}$ and $n_{\mu}$ being two null vectors,
\bqn
\lb{eq6}
l_{\mu} &=& \delta^{0}_{\mu},\;\;\;
n_{\mu} = \frac{1}{2}f(v,r)
\delta^{0}_{\mu} + \delta^{1}_{\mu},\nb\\
l_{\lambda}l^{\lambda} &=& n_{\lambda}n^{\lambda} = 0, \;\;
l_{\lambda}n^{\lambda} = - 1.
\eqn
The part $ T^{(n)}_{\mu\nu}$
of the EMT, can be considered as the
component of the matter field that moves along the null hypersurfaces $v
= Const.$ In particular, when $\rho = P = 0$, the solutions reduce to
the the N-dimensional
Vaidya solutions with $m = m(v)$  \cite{jai99,Chat}. 
Therefore, for the general case we
consider the EMT of Eq.(\ref{eq3}) as a generalization of the Vaidya
solutions in N-dimensional spacetimes.

Projecting the EMT of Eq.(\ref{eq3}) to the orthonormal basis, defined by
the unit vectors,
\bqn
\nb
E_{(0)}^{\mu}& = & \frac{l_{\mu} + n_{\mu}}{\sqrt{2}}, \\
\nb
E_{(1)}^{\mu}& = & \frac{l_{\mu} - n_{\mu}}{\sqrt{2}},\\
\lb{eq7}
E_{(2)}^{\mu}& = & \frac{1}{r}\delta^{\mu}_{2},\\
\nb
E_{(i)}^{\mu}& = & \frac{1}{r
\left[\prod_{l=2}^{i-1}\sin^{2}\left(\theta^{l}\right)
\right]} \delta^{\mu}_{i}, \; \; \; \; \; 
(i= 3,4, ..N-1),
\eqn
we find that
$T_{(a)(b)}
\equiv
e^\mu_{(a)} e^\nu_{(b)}
T_{\mu\nu}
$  takes the form
\bq
\lb{eq8}
\left[T_{(a)(b)}\right] = \left[\matrix{
\frac{\mu}{2} + \rho& \frac{\mu}{2}& 0 & 0 & ... &0 \cr
\frac{\mu}{2} & \frac{\mu}{2} - \rho & 0 & 0  & ...& 0\cr
0 & 0&  P &0 & ... & 0\cr
0 & 0& 0& P & ... & 0\cr
.. & .& . & . & ... & .\cr
.. & .& . & . & ... & .\cr
.. & .& . & . & ... & .\cr
0 & 0 &0 & 0 & ...& P \cr}\right],
\eq
which belongs to the Type II fluids defined in \cite{HE1973}.
The null vector $\l^{\mu}$ is the
double null eigenvector of the EMT. 
Physically acceptable solutions
have to satisfy several energy conditions  
\cite{HE1973}.

Following  \cite{VRW2000},
let us define the mass function $m(v,r)$
as
\bq
\lb{defm}
m(v, r) \equiv \frac{B_N}{2}{r^{N-3}}(1 +
r_{,\alpha}r_{,\beta} g^{\alpha \beta})
\eq
\noindent where
$()_{,\alpha}$ $=$
$\partial()/\partial x^\alpha$ and $B_N$
is a constant defined by
\bq
\lb{defbn}
B_N = \frac{\kappa \Gamma 
\left( \frac{N-1}{2}\right)}{2 (N-2) \pi^{(N-1)/2} } .
\eq
In the present case, it can be shown  that $m(v,r)$ takes the form
\bq
m(v,r) \; = \; \frac{B_N}{2}{r^{N-3}} \left[ 1 - f(r,v) \right]   ,
\eq
\noi or inversely,
\bq
\lb{deff}
f(v,r) \; = \; 1 - 2 \frac{ m(v,r)}{B_N r^{N-3}} .
\eq
Clearly, from the above definition, 
we find that the apparent horizon is located
on the hyppersurface defined by
\bq
\lb{defah}
r_{,\alpha}r_{,\beta} g^{\alpha \beta} =0.
\eq
\noindent In terms of $m(r,v)$, the
Kretschmann Scalar reads 
\bqn
\nb
{\cal{R}}
& \equiv & R_{\alpha \beta\gamma\delta}
R^{\alpha \beta\gamma\delta}
\\ \nb 
& =& \frac{6}{B_N^2 r^{2(N-1)}}
\left\{ \left[ (N-3)^2 (N-2)^2 + 4 \sum_{k=1}^{N-3} k \right] m^2
+ 4 r^4 m''^2 \right.
\\ \lb{defkm}  & &{\phantom{{ \frac{6}{B_N^2 r^{2(N-1)}}}}}
\; \; \; + 4 (N-3)^2 
\left[(N-2)r m m'' + r^2 m'^2 \right]
\\ \nb & &
\left. {\phantom{{ \frac{6}{B_N^2 r^{2(N-1)}}}}}
\; \; \; \; 
- 2 (N-3)\left[
(N-2)m + 2 r m' \right] r^2 m'' \right\} ,
\eqn
\noi while Eqs. (\ref{eq5}) takes the form
\bqn
\nb
\mu  & = &  \frac{N-2}{\kappa
B_N}\frac{ \dot{m}(v, r)}{ r^{N-2}} \\
\lb{murhopn}
\rho & = &   \frac{N-2}{\kappa B_N} \frac{m'}{r^{N-2}}\,  \\
\nb
P &= &- \frac{1}{\kappa B_N} \frac{ m''(v, r)}{r^{N-3}}.
\eqn

Without loss of generality, we expand $m(v,r)$ in the powers of $r$,
\bq
\lb{mgenn}
m(v, r) = \sum^{+ \infty}_{k = - \infty} a_{k}(v) r^{k},
\eq
where $a_{k}(v)$ are arbitrary 
functions of $v$ only.  Note that the sum of the
above expression should be understood as an integral, when the
``spectrum'' index $k$ is continuous. With this definition,
from Eq. (\ref{murhopn}) we have
\bqn
\lb{expr}
\mu &=& \frac{(N-2)}{\kappa B_N}
\sum^{+ \infty}_{k = - \infty}
{\dot{a}_{k}(v)r^{ k -N + 2}},\;\;\;
\rho = \frac{(N-2)}{\kappa B_N}
\sum^{+ \infty}_{k = - \infty}
{k{a}_{k}(v)r^{k - N + 1}},\nb\\
P &=& - \frac{1}{\kappa B_N}
\sum^{+ \infty}_{k = - \infty}
{k(k-1)a_{k}(v)r^{k-N +1}}.
\eqn
Once we have the general solutions, let us consider
some particular cases.

\noi {\bf  i) The Null Fluid Solution}: Let us consider first the 
null dust case, defined by the conditions
\bq
\lb{ndc}
P \; = \; 0 \; = \; \rho 
\eq
which in terms of (\ref{mgenn}) can be written as
\bq
\lb{expmnd}
a_{k}(v) = \cases{
m(v) ,& $k = 0$,\cr
0, & $k \not= 0$\cr}.
\eq
\noindent These are the N-Dimensional 
Vaydia solutions \cite{jai99,Chat},
in which the three energy conditions,
weak, strong and dominant,
all reduce to $\mu \ge 0$. For
these solutions, the Kretschmann Scalar (\ref{defkm})
reads
\bq
{\cal{R}} = \frac{6}{B_n^2}
\left\{(N-3)^2
(N-2)^2 + 4 \sum_{k=1}^{N-3} k  
\right\} \frac{h^2(v)}{r^{2(N-1)}},
\eq 
\noi which is singular when $r=0$.

\noi {\bf ii) The de Sitter and the Anti-de Sitter solutions}:
Another simple case is the consideration of the vacuum equations
in the presence of the cosmological constant $\Lambda$.
The solutions of the Einstein equations for this case
are given by
\bq
\lb{explamb}
a_{l}(v) = \cases{
\frac{B_N \Lambda}{(N-2)(N-1)},& $k = N-1$,\cr
0, & $k \not= N-1 $.\cr}
\eq
\noi {\bf iii) The Charged Vaidya Solutions}:
The consideration of the eletromagnetic field  $F_{\mu\nu}$, given by
\bq
\lb{eq20}
F_{\mu\nu} = \frac{q(v)}{r^{N - 2}}(\delta^{0}_{\mu}\delta^{1}_{\nu}
- \delta^{1}_{\mu}\delta^{0}_{\nu}),
\eq
\noindent yields
\bq
\lb{expmq}
a_{k}(v) = \cases{
f(v), & $k = 0$,\cr
- \frac{ B_N q^{2}(v)}{(N-2)(N-3)},& $k = - (N-3)$,\cr
0, & $k \not= 0, -(N-3) $.\cr}
\eq
\noindent The quantities (\ref{murhopn}) are now
\bqn
\rho & = & P = \frac{q^2(v)}{\kappa r^{2(N-2)}} , \\ \nb
\mu  & = & \frac{1}{\kappa B_N (N-3)r^{2N-5}} 
\left[ 
(N-2)(N-3)r^{N-3} \dot f(v) - {2}B_N q(v) \dot q(v)
\right] .
\eqn
The condition $\mu \geq 0 $ gives the main 
restriction on the choice of the functions $f(v)$
and $q(v)$. In particular, if $df/dq > 0$, we can see that
there always exists a critical radius $r_c$, 

\bq
\lb{emrc}
r_c = \left[
\left(\frac{2 B_N}{(N-2)(N-3)} \right)
\frac{q \dot q}{\dot f}
\right]^{\frac{1}{N-3}}.
\eq
\noindent When $r < r_c$, we have $\mu < 0$, 
and the energy conditions are always violated.

\noindent In this case, the apparent horizon is given by
\bq
\lb{ahef}
r_{ah} \; = \left\{
B_N^{-1}\left[
f \pm \sqrt{\frac{(N-2)(N-3)f^2(v) -2 q(v)^2}{(N-2)(N-3)}}
\right] 
\right\}^{\frac{1}{N-3}} .
\eq

It can be show that the corresponding 
Kretchmann Scalar is proportional to \( r^{-4(N-2)}\).

\noi {\bf iv) The Solutions with Linear Equation of State}: 
We will consider now the case for the  equation of state
\bq
\lb{sel}
P = \alpha {\rho} ,
\eq
\noindent where $\alpha$ is a constant.
Solving the equations (\ref{murhopn}) and (\ref{sel})
we found two irreductible cases: 
$\alpha \neq 1/(N-2)$ and $\alpha = 1/(N-2)$.

When $\alpha \neq 1/(N-2)$, we have
\bq
\lb{eq22}
a_{k}(v) = \cases{
h_1(v), &  $k = 0$,\cr
- \frac{h_0(v)}{ [\alpha(N-2) - 1]},&  $k =1-
\alpha (N-2)$,\cr
0,  &  $k \neq 0, 1 - \alpha (N-2) $.\cr}
\eq
Then the apparent horizon is given by 
\bq
\lb{rahsel}
r^{\alpha(N-2)+(N-4)}
- \frac{2 h_1(v) }{B_N} r^{\alpha(N-2) -1}
+ \frac{4h_0}{B_N[ \alpha(N-2) -1]} = 0     .
\eq
\noindent The corresponding Kretschmann scalar is proportional to
\( r^{-2[(N-2)(\alpha+1)]}\), and 
singularities at $r=0$ 
will occur for \( \alpha \geq -1
\).
>From Eqs.(\ref{murhopn}) we find
\bqn
P\, = \, \alpha \rho  &= & \alpha 
\frac{(N-2)}{\kappa B_N}
\frac{h_0(v) }{ r^{(N-2)(\alpha+1)} } ,
\nb \\
\mu  & = &
\frac{(N-2)}{[\alpha(N-2)-1]B_N
r^{(N-2)(\alpha+1) -1} } \; \; \times \\ \nb
& & \; \; \; \; \; \; \; \; 
\left\{\alpha[(N-2)-1]B_N
r^{(N-2)(\alpha+1) -1} \dot h_1(v) - \dot h_0 (v) 
\right\} .
\eqn 
When \(\alpha = 1/(N-2)\) the solution 
for Eqs.  (\ref{murhopn}) and (\ref{sel}) is given by
\bq
\lb{eq22a}
a_{k}(v) = \cases{
h_1(v) + h_2(v) \ln[r], & $k = 0$, \cr
0,  &  $k \neq 0 $,\cr}
\eq
\noi while Eqs. (\ref{murhopn}) then yield 
\bqn
P =\frac{1}{N-2}\rho & = & \frac{h_2(v)}{\kappa B_N r^{N-1}} 
\\ \nb
\mu  & = &
\frac{(N-2)}{\kappa B_N
r^{(N-2)} }
\left\{
\dot h_1(v) - \dot h_2(v)\ln[r]  
\right\} 
\eqn 
\noi and the corresponding apparent horizon is given by
\bq
\frac{B_N}{2} r^{N-3} - h_1(v) - h_2(v) \ln[r] \, = \, 0.
\eq

\noi v) {\bf The Solutions with the Polytropic Equation of State:}
A polytropic equation of state is defined as
\bq
\lb{se}
P = \alpha {\rho}^\beta 
\eq
\noindent where $\alpha$ and $\beta$ are constants. 

Substituting Eqs. (\ref{se}) into  Eqs. (\ref{murhopn})
we find that for $\beta \neq 1$ the function $m(v,r)$
is given by
\bq
\lb{msenl}
m(v,r) \; = \; 
\int \left\{ [1- \beta ] h_0(v)
- \alpha \left[ \frac{B_N}{N-2}  r^{N-2} \right]^{1-\beta}
\right\}^{\frac{1}{1-\beta}} dr
+ h_1(v) ,
\eq
\noi while  Eqs. (\ref{murhopn}) now read
\bqn
\nb
P & = & \alpha \rho^\beta  \, =  \, \frac{\alpha}{\kappa^{\beta-1}}
\left[ 
\frac{(N-2)}{\kappa B_N r^{N-2}} \right]^{\beta}
\left\{  [ 1 - \beta ]   h_0 (v)
- \alpha \left[ \frac{B_N}{N-2}  r^{N-2} \right]^{1-\beta}
\right\}^{\frac{\beta }{1-\beta}} ,
\\ 
\mu  & = & \frac{\beta (N-2) \dot h(v)}{\kappa B_N r^{N-2}} 
\int \left\{ [1-\beta] h_0 (v)
- \alpha \left[ \frac{B_N}{N-2}  r^{N-2} \right]^{1-\beta}
\right\}^{\frac{-\beta}{1-\beta}} dr + \nb \\
& & \; \; \; \; \; 
+ \dot h_1(v) .
\eqn
The apparent horizon is given by the equation
\bqn
\nb
r_{ah}^{N-3} & -& \frac{2}{B_N} 
\int \left\{ [1-\beta] h_0(v)
-  \alpha \left[ \frac{B_N}{N-2}  r^{N-2}_{AH} \right]^{1-\beta}
+
\right\}^{\frac{1}{1-\beta}} dr  +\\
&- & \frac{2}{B_N}  h_1(v)
=0.
\eqn
\section{N-Dimensional Vaidya Solutions in Double-Null Coordinates}
In double null coordinates, we can write the spherically symmetric
N-dimensional spacetime as \cite{jai99}
\bq
\lb{dn1}
ds^2 =2 f(u,v) du dv - r^2(u,v) d\Omega^{2}_{N-2},
\eq
\noi where $u$ and $v$ are the double null coordinates and
$d\Omega^{2}_{N-2}, $ is the line element
on the unit (N-2)-sphere, defined by (\ref{eq1a}).

The EMT  for a null dust fluid is given now by
\bq
\lb{dn2}
T_{\mu\nu} = \mu(u,v) \delta^v_{\mu}\delta^v_{\nu} .
\eq
Then the Einstein's field equations take the form
\bqn
\lb{dn3a}
\frac{(N-2)}{r f } 
\left( f r_{,u u} -{f_{,u}} r_{,u}\right) & = & 0 
\\
\lb{dn3b}
\frac{2}{f} \left[ r r_{,u v}
+(N-3){r_{,u} r_{,v}}\right] +(N-3) & = & 0 
\\
\lb{dn3c}
-\frac{(N-2)}{rf} \left( f r_{,vv} -{f_{,v}} r_{,v}\right) & = & 
\mu (u,v) 
\\
\lb{dn3d}
r \left(f{f_{,uv}} -{f_{,u} f_{,v}}\right) + (N-2)f^2r_{,u v}
& = & 0 .
\eqn
Integrating Eq. (\ref{dn3a}) with respect to \(u\), we 
obtain
\bq
\lb{dn4}
f \; = \; u_0 h_1 (v) r_{,u}
\eq
\noi where $u_0$ is an arbitrary constant and  
$h_1$ an arbitrary function. 

Differentiating  Eq. (\ref{dn3b}) with respect to $u$, considering 
Eq.(\ref{dn3a}), then integrating 
the result with respect to $u$,  we find
\bq 
\lb{dn5}
r^{N-2}\frac{r_{uv}}{f} + h_0(v) \; = \; 0 ,
\eq
\noi where $h_0(v)$ is another
arbitrary function. Setting now Eqs.
(\ref{dn4}) and 
(\ref{dn5}) in (\ref{dn3b})
we have
\bq
\lb{dn6}
r_{,v} \; = \; -{h_1} \left[
1 - \frac{2}{N-3} \frac{h_0}{r^{N-3}}
\right] ,
\eq
\noi where in writting the above expression, we set $u_0=2$.
Using this definition,
Differentiating Eq. (\ref{dn3b}) with relation to $v$,
and considering Eqs. (\ref{dn3c}), (\ref{dn3d}), (\ref{dn5}),
and (\ref{dn6}) we have
\bq
\lb{dn8}
\mu \; =\; 2
\frac{N-2}{N-3} 
\frac{{h_0}_{,v}}{r^{N-2}} .
\eq
Equations (\ref{dn4}), (\ref{dn5}), (\ref{dn6})
and (\ref{dn8}) are, respectivaly, the N-dimensional
generalization of equations (11), (9), (12)
and (13) in \cite{WL86}.

>From Eq.(\ref{defm}), we find
\bq
\lb{dn9}
m (u,v ) \; = \; \frac{B_N}{2 } r^{N-3}
\left[
1 + 2 \frac{r_{,u}r_{,v}}{f}
\right]
\; = \; \frac{B_N}{N-3} h_0(u,v)
\eq
\noi or inversely,
\bq
\lb{dn10a}
h_0(u,v) \; = \; \frac{N-3}{B_N} m(u,v) .
\eq
Now let us consider the case where 
\bq
\lb{dnm}
m \; = \; \lambda v^{N-3},
\eq
\noi with $\lambda >0$, so the solutions have homothetic
self-similarity \cite{blrw01}. Then, from Eq. (\ref{dn6}) we find
\bq
\lb{dn14}
\int \frac{x ^{N-3} dx}{x^{N-2} - \frac{x^{N-3}}{2 } 
+\lambda }  +ln[v] = h_2(u)
\eq
\noi where  
\bq
\lb{dnx}
x\; = \; \frac{r}{v} 
\eq
\noi is the self-similar 
variable and $h_2$ an arbitrary function. From 
(\ref{dn4}) and (\ref{dn14}), and setting $h_2(u) = - u$,
we find 
\bq
\lb{dn15a}
f \; = \;  \frac{r}{x^{N-2}}
\left[
x^{N-2} - \frac{1}{2}x^{N-3} + \lambda 
\right]  \; \equiv \; \frac{r}{x^{N-2}} F (x) .
\eq

Depending on the specific value of $\lambda$, it can be
show that there are three distinguishible cases:
i)$\lambda < \lambda_c (N)$;
ii) $\lambda = \lambda_c (N)$; and 
iii) $\lambda > \lambda_c (N)$ [cf. Fig.1].
When $\lambda < \lambda_c$, $F(x)=0$ in general
has two real roots, says $x_1$ and $x_2$. 
Without loss of generality, we assume
$x_2$ $>$ $x_1$. 
Similar to the case $N = 4$, the one with $x= x_2$
corresponds to the Cauchy horizon and the corresponding Penrose diagram
is given by Fig 2(a). Thus, in this case,
the collapse forms naked singularities.
When $\lambda = \lambda_c$, the two real roots, $x_1$
and $x_2$ degenerates to one, $x_c$ $=$ $x_1$ $=$ $x_2$,
and we have ${F}_{,x} (x_c) = 0$, i.e,
\bq
\lb{dnxc}
x_c - \frac{N-3}{2(N-4)} \,= \, 0 .
\eq
\noi Substituting Eq.(\ref{dnxc}) into the equation 
$F(x_c) =0$, we find that
\bq
\lb{dnlc}
\lambda_c \; = \; \frac{\left(
N-3\right)^{N-3}}{ \left[2 (N-2) \right]^{N-2}} .
\eq
\noi The corresponding Penrose diagram is 
given by Fig. 2(b).
It  is remarkable
to note that $\lambda_c$
depends only on the spacetime dimension $N$.

When $\lambda > \lambda_c$,
we can see that the equation $ F(x) = 0$ has no real roots
and the function $f(u,v)$ is strictly positive,
$f(u,v) > 0$.
Similar to the four dimensional case, it can be show that now the
collapse always forms black holes and the corresponding Penrose 
diagram is given by Fig. 2(c).

\section{Conclusions }
In this paper, we have first
generalized the solutions found
by Wang and Wu \cite{Wang1998}
in four-dimensional spherically symmetric spacetimes
to N-dimensional spacetimes
and found that the most known solutions belong to the
large class of solutions presented there. 
Then, we have restricted our attention
to the ingoing Vaidya solutions, which can be interpreted
as representing gravitational collapse of null dust 
fluid in N-Dimensional
spacetime. To study the global structure of the
corresponding spacetimes, following Waugh and Lake
\cite{WL86}, we found the explicit close form
of the metric in double null coordinates for the solutions
with homothetic self-similarity \cite{blrw01}
and shown explicitly that when $\lambda > \lambda_c$,
the collapse always forms black holes, and
when $\lambda < \lambda_c$, it always
forms naked singularities. The critical value $ \lambda_c$
has been also found and it is remarkable
that it depends only on the space dimension
N,
$$
\lambda_c \; = \;
\frac{(N-3)^{N-3}}{[2(N-2)]^{N-2}} .
$$
\noi When N=4 we have  $\lambda_c=1/16$,
which is exactly the value found first by Papaetrou
\cite{P1984}
and later confirmed by several authors.

{\em Note added}: After this paper has been submitted for publication, a
preprint appeared in {\tt xxx.lanl.gov } \cite{GD01},  in which the
self-similar Vaidya solutions were also studied, but by a different method.  
Moreover,  we have been lately also informed that similar considerations of
null dust solutions in higher dimensional spacetimes were given in \cite{PD99}.

\section*{Acknowledgments}

The author would like very much to thank A. Z. Wang for valuable suggestions
and discussions, and CNPq for  financial assistance.
%%%%%%%%%%%%%%%%%%%%%%%%%%%%%%%%%%%%%%%%%%%%%%%%%%%%%%%%%%%%%%%%%%%%%%%%%%%

\newpage

\begin{figure}[htbp]
\begin{center}
\leavevmode
\epsfig{file=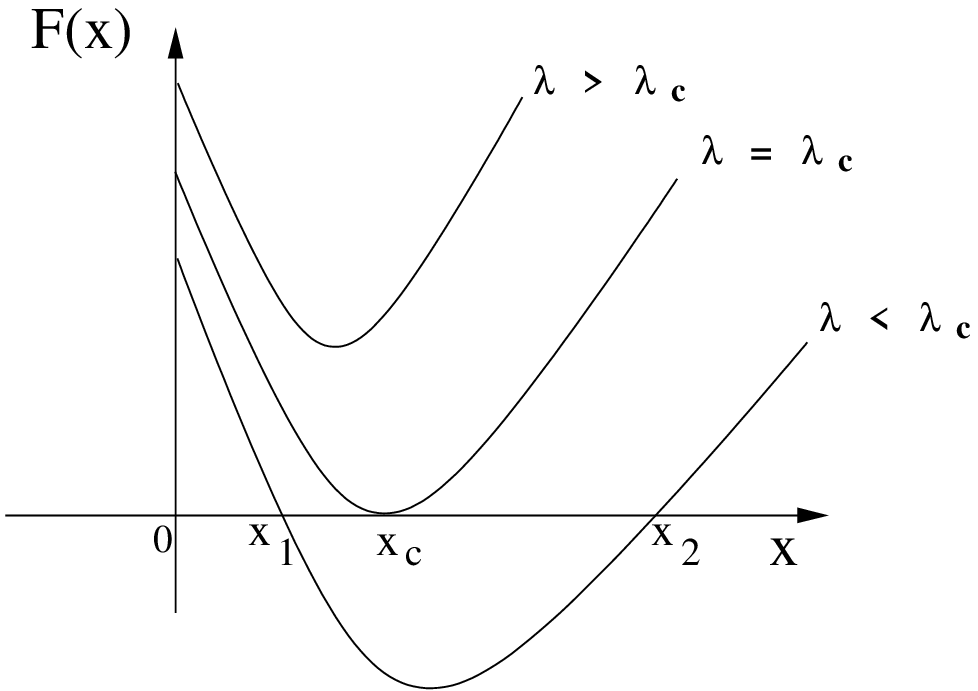,width=0.4\textwidth,angle=0}
\caption{{\bf Fig. 1} The qualitative behavior
of the function F(x) defined by Eq. (\ref{dn15a})  
in the text.}
\label{fig1}
\end{center}
\end{figure}

\begin{figure}[htbp]
\begin{center}
\leavevmode
\epsfig{file=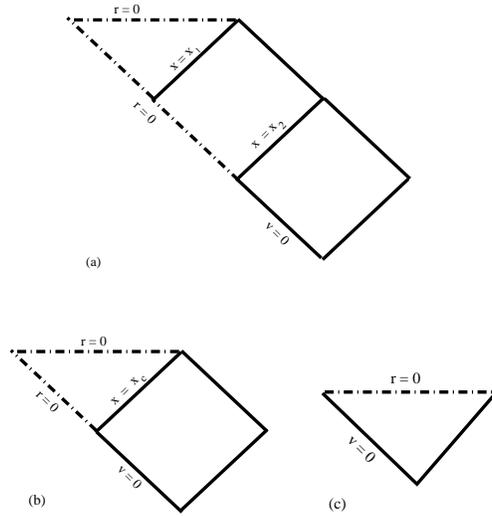,width=0.4\textwidth,angle=0}
\caption{{\bf Fig. 2}  Penrose's diagrams:  (a) the  $\lambda < \lambda_c$
case; (b) the case  $\lambda = \lambda_c$; and
(c) the $\lambda > \lambda_c$ case.}
\label{fig2}
\end{center}
\end{figure}

\end{document}